\documentclass[longauth]{aa} 
\usepackage{graphicx}
\usepackage{txfonts}
\usepackage{natbib}
\begin{document}
   \title{Radio emission of highly inclined cosmic ray air showers 
measured with LOPES}


\author{J.~Petrovic\inst{1}
\and W.D.~Apel\inst{2}
\and T.~Asch\inst{3}
\and F.~Badea\inst{2}
\and L.~B\"ahren\inst{4}
\and K.~Bekk\inst{2}
\and A.~Bercuci\inst{5}
\and M.~Bertaina\inst{6}
\and P.L.~Biermann\inst{7}
\and J.~Bl\"umer\inst{2,8}
\and H.~Bozdog\inst{2}
\and I.M.~Brancus\inst{5}
\and M.~Br\"uggemann\inst{9}
\and P.~Buchholz\inst{9}
\and S.~Buitink\inst{1}
\and H.~Butcher\inst{4}
\and A.~Chiavassa\inst{6}
\and F.~Cossavella\inst{8}
\and K.~Daumiller\inst{2}
\and F.~Di Pierro\inst{6}
\and P.~Doll\inst{2}
\and R.~Engel\inst{2}
\and H.~Falcke\inst{1,4,7}
\and H.~Gemmeke\inst{3}
\and P.L.~Ghia\inst{10}
\and R.~Glasstetter\inst{11}
\and C.~Grupen\inst{9}
\and A.~Haungs\inst{2}
\and D.~Heck\inst{2}
\and J.R.~H\"orandel\inst{8}
\and A.~Horneffer\inst{1}
\and T.~Huege\inst{2}
\and K.-H.~Kampert\inst{11}
\and Y.~Kolotaev\inst{9}
\and O.~Kr\"omer\inst{3}
\and J.~Kuijpers\inst{1}
\and S.~Lafebre\inst{1}
\and H.J.~Mathes\inst{2}
\and H.J.~Mayer\inst{2}
\and C.~Meurer\inst{2}
\and J.~Milke\inst{2}
\and B.~Mitrica\inst{5}
\and C.~Morello\inst{10}
\and G.~Navarra\inst{6}
\and S.~Nehls\inst{2}
\and A.~Nigl\inst{1}
\and R.~Obenland\inst{2}
\and J.~Oehlschl\"ager\inst{2}
\and S.~Ostapchenko\inst{2}
\and S.~Over\inst{9}
\and M.~Petcu\inst{5}
\and T.~Pierog\inst{2}
\and S.~Plewnia\inst{2}
\and H.~Rebel\inst{2}
\and A.~Risse\inst{13}
\and M.~Roth\inst{2}
\and H.~Schieler\inst{2}
\and O.~Sima\inst{5}
\and K.~Singh\inst{1}
\and M.~St\"umpert\inst{8}
\and G.~Toma\inst{5}
\and G.C.~Trinchero\inst{10}
\and H.~Ulrich\inst{2}
\and J.~van~Buren\inst{2}
\and W.~Walkowiak\inst{9}
\and A.~Weindl\inst{2}
\and J.~Wochele\inst{2}
\and J.~Zabierowski\inst{13}
\and J.A.~Zensus\inst{7}
\and D.~Zimmermann\inst{9}
}

\authorrunning{J. Petrovic et al.}
\titlerunning{Radio emission of highly inclined showers
measured with LOPES}

\institute{Radboud University Nijmegen, Department of Astrophysics, IMAPP,
           P.O. Box 9010, 6500 GL Nijmegen, The~Netherlands\\
              \email{petrovic@astro.ru.nl}\\
             \and
            Institut\ f\"ur Kernphysik, Forschungszentrum Karlsruhe,
            76021~Karlsruhe, Germany\\
           \and
            Institut f\"ur Prozessverarb. und Elektr., Forschungszentrum Karlsruhe,
            76021~Karlsruhe, Germany\\
           \and
           ASTRON, 7990 AA Dwingeloo, The Netherlands \\
           \and
            National Institute of Physics and Nuclear Engineering,
            7690~Bucharest, Romania\\
            \and
            Dipartimento di Fisica Generale dell'Universit{\`a},
             10125 Torino, Italy\\
\and
Max-Planck-Institut f\"ur Radioastronomie,
53010 Bonn, Germany \\
\and
Institut f\"ur Experimentelle Kernphysik,
Universit\"at Karlsruhe, 76021 Karlsruhe, Germany\\
\and
Fachbereich Physik, Universit\"at Siegen, 57068 Siegen,
Germany \\
\and
Istituto di Fisica dello Spazio Interplanetario, INAF,
10133 Torino, Italy \\
\and
Fachbereich Physik, Universit\"at Wuppertal, 42097
Wuppertal, Germany \\
\and
Radioastronomisches Institut der Universit\"at Bonn,
53121 Bonn, Germany \\
\and
Soltan Institute for Nuclear Studies, 90950~Lodz,
Poland\\
}

   \offprints{petrovic@astro.ru.nl}

   \date{}

 
  \abstract
   {}
   {The capability of radio antenna arrays to measure cosmic 
ray air showers with very large zenith angles is explored.
This is important, since a possible neutrino detection has to fulfill two 
requirements. First: antennas should be able to detect very inclined 
cosmic ray 
air showers, and second: it should be possible to estimate the distance to 
the shower maximum, since neutrinos are most likely to 
travel far through the Earth's atmosphere without interaction and induce
air showers close to the ground.} 
   {LOPES (LOFAR PrototypE 
Station; LOFAR - LOw Frequency ARray), an array of dipole antennas, is 
used for the detection of inclined cosmic ray air showers. 
LOPES is co-located and triggered by the KASCADE 
(KArlsruhe Shower Core and Array DEtector) experiment, which also provides 
information on 
air shower properties such as electron and muon numbers on the ground, as 
well as the arrival direction.
Radio emission of nearly vertical cosmic ray air showers has been 
detected by LOPES.}
   {LOPES-10 (the first phase of LOPES, consisting of 10 antennas) 
detected a significant number of cosmic ray air 
showers with a zenith angle larger than 50$^{\circ}$, 
and many of these have very high radio field strengths.
The most inclined event that has been detected with LOPES-10 has
a zenith angle of almost 80$^{\circ}$. 
This is proof that the new technique is also applicable for 
cosmic ray air showers with high inclinations, which in the case that 
they are initiated close to the ground, can be a signature of neutrino events.}
   {Our results indicate that arrays of simple radio antennas 
can be used 
for the detection of highly inclined air showers,
which might be triggered by neutrinos. In addition, 
we found that the radio pulse height (normalized with the muon
number) for highly inclined events  
increases with the geomagnetic 
angle, which confirms the geomagnetic origin of 
radio emission in cosmic ray air showers.}


   \maketitle
%

\section{Introduction}
When cosmic rays interact with particles in the Earth's atmosphere, they produce
a shower of elementary particles propagating towards the ground almost at
the speed of light. The first suggestion that these air showers can produce radio emission was given by \citet{askaryan} 
based on a charge-excess mechanism. In some experiments during the sixties and seventies, 
coincidences between radio pulses 
and cosmic ray events were reported \citep{jelley,allan}. Recently, \citet{2003APh....19..477F} suggested that the 
mechanism for a radio emission of air showers might be geosynchrotron emission: 
secondary electrons and positrons produced in the particle cascade are deflected
in the Earth's magnetic field producing radiation that is relativistically beamed in the forward direction. 
They suggested that this radiation can be observed in the radio 
domain with the new generation of digital radio telescopes
like LOFAR \citep[LOw Frequency ARray,][]{2003NewAR..47..405R} and 
SKA \citep[Square Kilometer Array,][]{2005AN....326..608B}.
The shower in its densest region has a thickness that is smaller than the 
wavelength of radio emission below 
100 MHz (around a few meters). The emission is then coherent and the signal is amplified.
The fact that air showers can be observed in radio frequencies was 
recently confirmed by the first results of the LOPES 
\citep{2005Natur.435..313F} and CODALEMA experiments 
\citep{2004astro.ph..4240A}.

An air shower develops in three stages: an initial rapid build-up due to the cascade process, 
culminating in a particle number maximum, moving with relativistic speed.
At the maximum, ionization losses of electrons and positrons roughly equal 
their $\gamma$-ray production 
through bremsstrahlung (at a critical energy of about 80 MeV in air).
The third phase is a gradual decay as the electrons further lose energy through ionization.


Probably the most important task for radio and particle detectors in 
cosmic ray physics is finding  the composition and primary energy 
spectrum of cosmic rays, especially at ultra-high energies (UHECR). 
Currently, it is believed that cosmic rays (protons and nuclei) up to about 10$^{15}$ eV gain their energy in 
acceleration by shocks from supernova explosions \citep[see, for 
example,][]{2000RvMP...72..689N}.
For cosmic rays with even higher energies it has been suggested that acceleration can happen in 
the interaction with multiple supernova remnants \citep{1991aame.conf..273I}, galactic wind \citep{2004A&A...417..807V}, 
or extragalactic sources like
jets of gamma-ray bursts \citep{1995PhRvL..75..386W,1995ApJ...453..883V} or
active galactic nuclei \citep{1987ApJ...322..643B,1993A&A...272..161R}. 
Another option is so-called top-down scenarios:
protons and neutrons, but also neutrinos and gamma-rays, are produced from quark and gluon fragmentation of 
heavy exotic particles formed in the early Universe
\citep{hill,1983ICRC....2..393S}.
The highest energy cosmic rays are expected to be suppressed by interaction with the cosmic
microwave background radiation, which is known as the Greisen-Zatsepin-Kuzmin (GZK) cut-off
\citep{greisen,zatsepin}.
The HiRes experiment \citep{2004PhRvL..92o1101A} measurements are in agreement with the 
predicted flux suppression of 
the highest energy cosmic rays. However, the results of the AGASA 
experiment \citep{1998PhRvL..81.1163T} show an excess
compared to the predicted GZK suppression \citep{2003PhLB..556....1B}.
The ultra-high energy cosmic ray flux is expected to be accompanied
by associated fluxes of gamma rays and neutrinos
from pion decays formed in the collision of protons with photons  
\citep[see, for example,][]{2005NJPh....7..130Z}.

To reconstruct the initial energy of a primary particle (by measuring the 
particles on the ground with an array of particle 
detectors) it is necessary to have a good estimate of the total number of 
muons and electrons/positrons during shower developments. Muons suffer 
insignificant losses 
in the Earth's atmosphere and 
almost all reach the ground, so they are detected by particle detectors like KASCADE \citep[KArlsruhe Shower Core and 
Array DEtector, see][]{2003NIMPA.513..490A}. On the other hand, 
electrons are absorbed proportionally to the shower path length 
through the atmosphere, and, in case of highly inclined showers, 
only a small 
fraction reaches the ground, but the total number of electrons is expected to be roughly proportional 
to the strength of the radio pulse detected with radio arrays such as 
LOPES (LOFAR PrototypE Station).  
The numbers of muons and electrons in the air shower can also reveal the nature of the primary particle.
Cosmic ray air showers triggered by nuclei or protons can be recognized by the strong hadronic component, 
and since muons travel through the atmosphere almost unaffected, a large 
number of muons can be detected on the ground
level. Air showers that are initiated by gamma rays are dominated by the 
electromagnetic component 
(electrons, positrons, and photons) while the number of muons is lower.

Neutrinos are not very likely to interact within the Earth's atmosphere
and produce air showers due to the very small cross section for interactions. 
The probability
for interaction increases with the path length through the atmosphere, 
so neutrinos that
are most likely to trigger air showers are those entering the atmosphere 
almost horizontally, possibly 
producing very inclined cosmic ray air showers that may be detected in 
the radio domain. Those showers may be triggered anywhere along the 
traveling path and also close to the ground
\citep{1998APh.....8..321C}. 
The cross sections of neutrinos increase with 
their energy as well, 
so the highest energy neutrinos are more likely to be detected, 
but their expected flux is extremely
small.
It is interesting to mention that \citet{1975Ap&SS..32..461B} suggested 
that muons created in cosmic ray air showers that started high in the
atmosphere can trigger showers close to the ground, and these could appear 
similar to the neutrino-induced ones. However, with radio observations, 
due to the low 
attenuation of radio waves,
it should be possible to see 
two showers in this case: the original one in which secondary muons are 
created and the secondary one, triggered by
one of those muons. 

Highly inclined showers are expected to be very well detectable in the radio domain
\citep{2004APh....22..103G,2005A&A...430..779H,2005APh....24..116H} and they can be very
interesting for neutrino searches.
In this paper we show that the LOPES-10 radio array (the first phase of 
LOPES, consisting of 10 antennas)
can detect highly inclined cosmic ray air showers which can, in principle, be the signature of primary neutrinos. 
The data is also used to check the dependency of detection efficiency on 
muon number, zenith, and azimuth angle.
Also, highly inclined events are convenient for further investigation 
of the previously found 
correlation \citep{2005Natur.435..313F} between radio pulse 
strength and geomagnetic angle, 
due to the larger range of geomagnetic angle values.
The paper is organized in the following way:
in Sect. 2 we briefly introduce the LOPES experiment. In Sect. 3 we 
present some highly inclined events 
detection statistics and show an example event with zenith angle 
of 50$^{\circ}$. In Sect. 4 we 
show some properties of inclined showers. In Sect. 5 we present events 
with zenith angles larger than
70$^{\circ}$. Our conclusions are given in Sect. 6.  


\section{The LOPES experiment}

LOPES \citep[LOfar PrototypE Station, see][]{2004SPIE.5500..129H} is a phased array of dipole antennas co-located with
the KASCADE experiment \citep[Karlsruhe Shower Core and 
Array Detector, see][]{2003NIMPA.513..490A}, which provides coincidence 
triggers for LOPES and well-calibrated 
information about air-shower properties, like electron number $N_e$, 
total muon number $N_{{\mu},tot}$, 
truncated muon number $N_{\mu}$ 
(muon number within 20-400 m of the shower core), azimuth, and zenith 
angle of the event.
Radio emission of cosmic ray air showers (40-80 MHz) has been detected by 
LOPES-10 in the frequency range from 40 to 80 MHz
as reported by \citet{2005Natur.435..313F}.

The setup of LOPES-10 consists of 10 inverted V antennas distributed over 
the
KASCADE array, which consists of 252 scintillation detector stations on a 
uniform grid with 13 m spacing, 
electronically organized in 16 independent clusters. Each LOPES antenna is centered between four KASCADE huts in the north-western 
part of the array. The antenna setup has a maximum baseline of 125 m. The analogue radio signal received by antennas is 
filtered to select a band from 43 to 76 MHz, giving a bandwidth of 33 MHz.

About 0.8 ms of data are read out whenever KASCADE produces a `large-event 
trigger`, which means that 
10 out of 16 KASCADE clusters have produced an internal trigger. The resulting trigger rate is about two to three per minute
(depending on how many clusters are operating and local atmosphere 
conditions, such as pressure).  
On average the trigger is delayed with respect to the shower by 1.8 $\mu$s, 
depending on the shower geometry relative to the KASCADE detector.

We relate the radio signal to shower properties using parameters provided
by the KASCADE data processing from the particle detectors. These
parameters are: location of the shower core, shower direction, energy deposited in the particle detectors, 
total number of electrons on the ground level $N_e$, and muon numbers ($N_{{\mu},tot}$, $N_{\mu}$).
The strength of the detected radio pulse is in units $\mu$V/m/MHz.
There is no absolute gain calibration yet. 
The positions of radio flashes are coincident with the direction of the 
shower axis derived from KASCADE data within the expected uncertainties 
(average offset 0.8$\pm$0.4$^{\circ}$).
Details about the LOPES-10 experiment and data processing are given by 
\citet{2004SPIE.5500..129H} and \citet{2006PhDT........1H}. 

\section{Radio detection of inclined events}

\begin{figure*}[t]
\begin{center}
\includegraphics*[width=0.7\columnwidth,angle=270,clip]{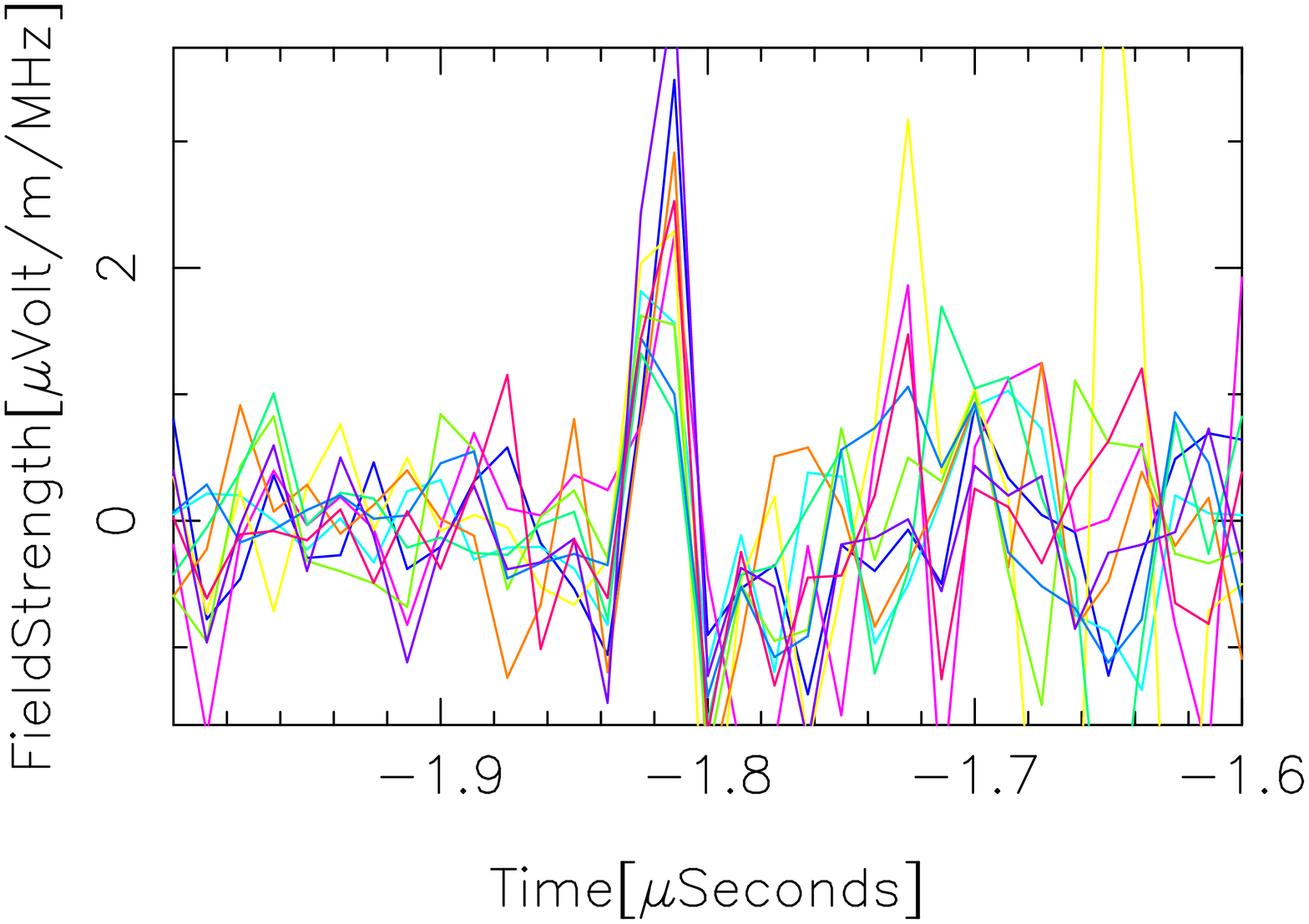}
\includegraphics*[width=0.7\columnwidth,angle=270,clip]{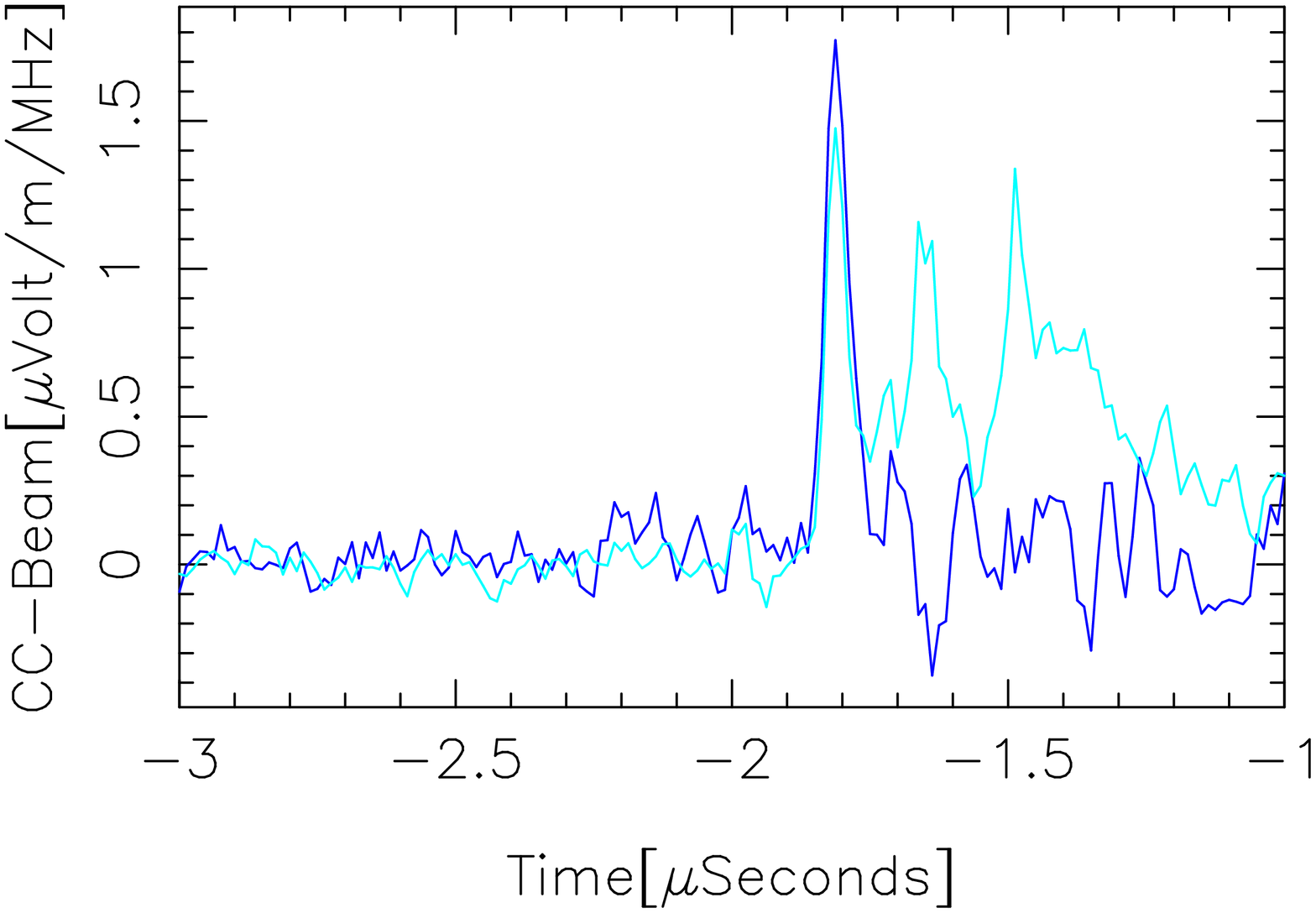}
\caption{\label{signal18-a}Left: The electric field as a function of time for 
LOPES-10 antennas for an event 
detected in March 2004 with a zenith angle of 53$^{\circ}$ and an azimuth angle of
54$^{\circ}$. Geomagnetic angle of this event is 70$^{\circ}$. Number of electrons is
$N_e$=1.5$\cdot$10$^6$ and total number of muons $N_{{\mu},tot}$=1.5$\cdot$10$^6$, reconstructed by KASCADE.
Right: Radio emission as a function of time after beam forming for the 
same event. The dark blue line represents the resulting cc-beam, 
the light blue line the power-beam. Note that the timescales 
on the panels are different.}
\end{center}
\end{figure*}

LOPES-10 received 2017 triggers from the KASCADE detector for 
events with a zenith angle larger than 50$^{\circ}$ in
seven months of taking data during the year 2004, 
with primary energy approximately in the range 
10$^{16}$-10$^{17}$ eV 
(which is roughly the threshold for the radio detection; 
the KASCADE detector is optimized for 10$^{15}$ eV). 
There were 1931 triggers for events with a zenith 
angle between 50$^{\circ}$ and 70$^{\circ}$ and, out of those, 
roughly 1 in 35 have been detected in the radio domain.
With the increase of zenith angle, the number of triggers 
decreases and there were only 86 events for
zenith angles larger than 70$^{\circ}$. Out of those events 
4 have been detected by LOPES-10.
This detection rate is significantly higher than the 
average detection rate for LOPES-10, 
which is roughly 1 event detected in the radio domain per 
1000 triggers received from
the KASCADE detector. The reason for this is that inclined 
showers that are detected by KASCADE
still have a significant number of particles reaching the 
ground, even after traveling
path lengths through the Earth's atmosphere 
several times longer than the vertical ones. In other words, inclined 
showers that
manage to reach the ground must have high energy primary particles, 
and they are above the threshold for radio detection.
It is also important to mention here that the radio signal of highly 
inclined events has a larger 'footprint' \citep{2005APh....24..116H} 
on the ground and 
due to this the effective collecting area of a radio antenna 
array becomes bigger.
This might be not so important for the small baseline arrays like 
LOPES-10 (around 100 m),
but it will be for larger arrays like LOFAR 
or Pierre Auger Observatory \citep{2000NuPhS..80C0811N}.

To reduce low energy air showers triggering KASCADE,
for the following detailed investigation of individual inclined cosmic 
rays, we introduce an additional condition: only showers with large 
densities in the muon counters are selected.
This narrows the sample to
51 events and around 40\% of those are detected by LOPES,
many with very large radio pulse heights.
The KASCADE detector is not optimized for large zenith angles, 
so the reconstruction of the electron and muon number is not accurate for 
all 51 events. However, it is important to mention here that shower 
direction and 
shower core position are always well reconstructed by KASCADE.

As an example, we show one of those 51 events (with accurate KASCADE
electron and muon number reconstruction), 
detected by LOPES-10 in March 
2004, with a zenith angle of
53$^{\circ}$ and an azimuth angle of 54$^{\circ}$
(an azimuth of 0$^{\circ}$ corresponds to a direction towards the 
north),
with roughly the same number of electrons and 
total number of muons $N_e$, $N_{{\mu},tot}$$\approx$1.5$\cdot$10$^6$. 
The angle between the shower axis and the Earth's magnetic 
field or so-called 
geomagnetic angle is 70$^{\circ}$. The energy of the primary 
particle is estimated to be around 
10$^{17}$ eV.

In Fig. \ref{signal18-a} (left panel) 
we show the electric field (defined as electric field strength divided 
by the LOPES bandwidth) as a function of 
time for each of ten LOPES-10 antennas, after being corrected
for instrumental and geometric time delays in the direction of the shower axis. 
The field is coherent at -1.825 $\mu$s, which is the arrival time of the 
shower. 
The incoherent noise afterwards is radio emission from photomultipliers 
and in this case is relatively weak,
indicating a lower energy deposit in the photomultipliers and a lower 
electron number on the ground than in the case of more vertical events,
due to the absorption along the shower's traveling path in the Earth's 
atmosphere.
The right panel of this figure shows the radio emission as a function 
of time after power-beam and cc(cross correlated)-beam forming.
For beam forming, for a given direction, signals are corrected for a time 
delay between the arrival of signals to each antenna and for antenna gain.
Then, the power beam is calculated as follows: 
the electric field of each antenna is squared, all is summed and 
normalized by 
the number of antennas, and the square root is taken.
The cross correlated beam is formed in 
the following way: 
electric field strengths of all 
possible two antenna combinations
are multiplied, summed, and normalized by the number of antennas, and 
finally, the square root is taken (while preserving the sign).
The power beam gives a peak if there is a lot of power in the antennas, 
independent of it being coherent or incoherent, while the cc-beam is
more sensitive to the coherence effects \citep[for 
more details see][]{2006PhDT........1H}.

\begin{figure}[h]
\begin{center}
\includegraphics*[width=0.8\columnwidth,clip]{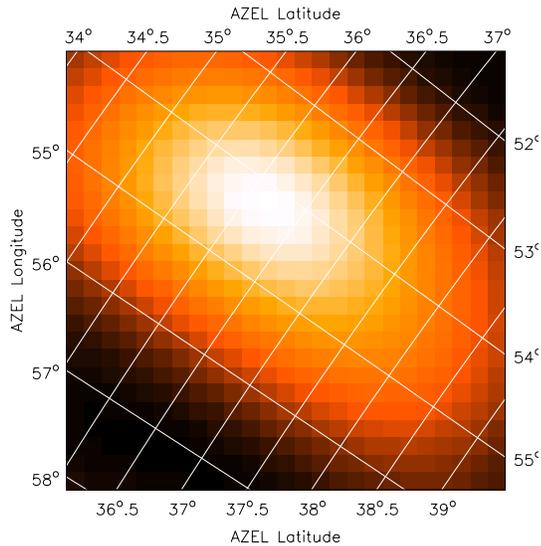}
\caption{\label{skymap}Radio map of the cosmic ray shower with a zenith angle of 53$^{\circ}$ 
detected in March 2004. 
Azimuth (AZEL longitude) and elevation (AZEL latitude) of the event are given on the axes.
The map is focused on a distance of 3000 m.}
\end{center}
\end{figure}

\begin{figure}[h]
\begin{center}
\includegraphics*[width=0.3\textwidth,clip]{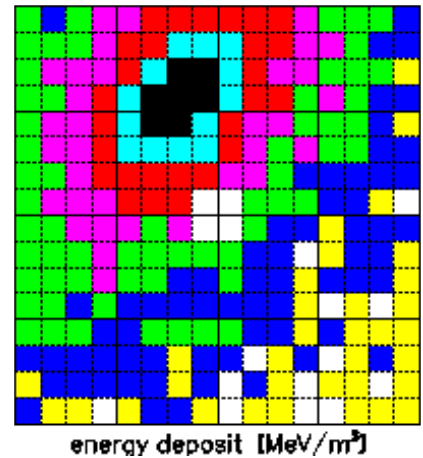}
\caption{\label{kascade}
Energy deposit of the  
cosmic ray shower with a zenith angle of 53$^{\circ}$ detected in March 2004 for e/$\gamma$ KASCADE detectors
(area 200x200 m$^2$, each small square represents one KASCADE station).
Dark blue color shows the energy deposit of 100 MeV/m$^2$, red color 
shows 1000 MeV/m$^2$. The maximum energy deposit
for this event is $\sim$4500 MeV/m$^2$.
}
\end{center}
\end{figure}

Figure \ref{skymap} is a radio map of the example event 
\citep{2005Natur.435..313F}. 
The air shower is the brightest point in the sky at the radio frequencies for
several tens of nanoseconds. 
Figure \ref{kascade} gives the energy deposit of the chosen cosmic ray 
shower over the KASCADE array with 
e/$\gamma$ detectors. 
We can see that the shower core falls within the KASCADE array and that
the maximum energy is deposited in the north-western part, within the 
LOPES-10 array. We  
can notice the elliptical shapes of the contours of the energy deposit, 
which is typical for inclined events.

We should also point out that the antenna gain 
(ratio of the power received/transmitted in a certain 
direction to the power that an isotropic emitter would radiate in that 
direction) 
decreases with an increase of zenith angle (the gain also depends on frequency
and azimuth), so
the sensitivity of LOPES-10 decreases towards the horizon. 
For example at fixed azimuth angle, the gain is around three times 
larger for a zenith angle of 
20$^{\circ}$ than for a zenith angle of 70$^{\circ}$.
We should keep in mind that due to this, on a map of an inclined event,
(for example Fig. \ref{skymap}), the maximum can be slightly shifted 
towards lower zenith angles.
Also, gain values for larger zenith angles are less certain.

\section{Properties of inclined showers}

\begin{figure}[h]
\begin{center}
\includegraphics*[width=\columnwidth,clip]{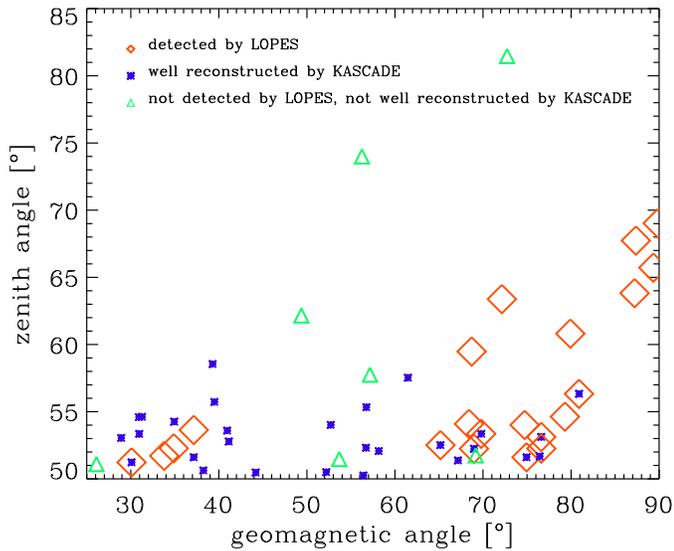}
\caption{\label{zenitgeo}Geomagnetic and zenith angle of inclined 
events from the year 2004. 
Rhombs: events detected by LOPES-10 in the radio domain,
stars: events with the reliable KASCADE reconstruction of 
muon and electron number, and triangles: events not detected by LOPES
and without well reconstructed muon and electron numbers by KASCADE.  
Radio detection efficiency is almost 100\% above a geomagnetic angle of 60$^{\circ}$.}
\end{center}
\end{figure}

\begin{figure}[h]
\begin{center}
\includegraphics*[width=\columnwidth,clip]{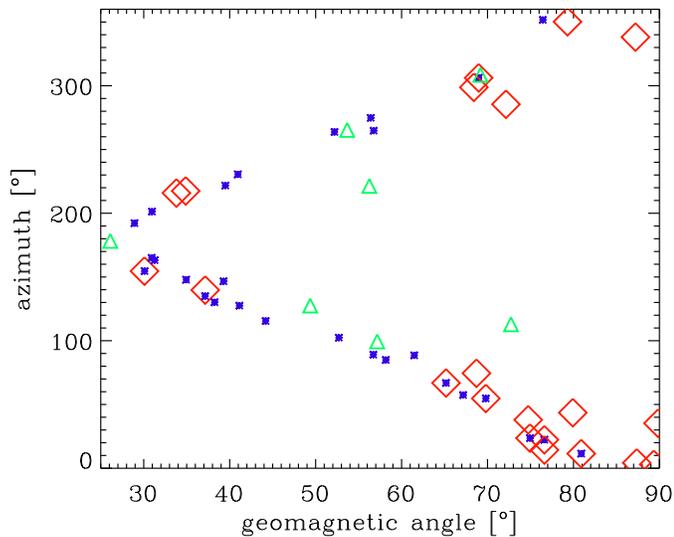}
\caption{\label{azimgeo}Geomagnetic angle and azimuth for the same 
set of events as in Fig. \ref{zenitgeo}. 
The meaning of symbols is the same as in Fig. \ref{zenitgeo}.
Note that there are more events detected by LOPES-10 in the north-south 
direction than in the east-west
direction.
This may be due to the antenna sensitivity, 
largest geomagnetic angles in the
direction of north or polarization effects. }
\end{center}
\end{figure}

Out of 51 events with a zenith angle larger than 50$^{\circ}$, 
two occurred in thunderstorm conditions,
and are taken out of the further analysis \citep{buitink}.
The remaining 49 events are shown in Fig. \ref{zenitgeo} and 
\ref{azimgeo}.
 
\begin{figure}[h]
\begin{center}
\includegraphics*[width=\columnwidth,clip]{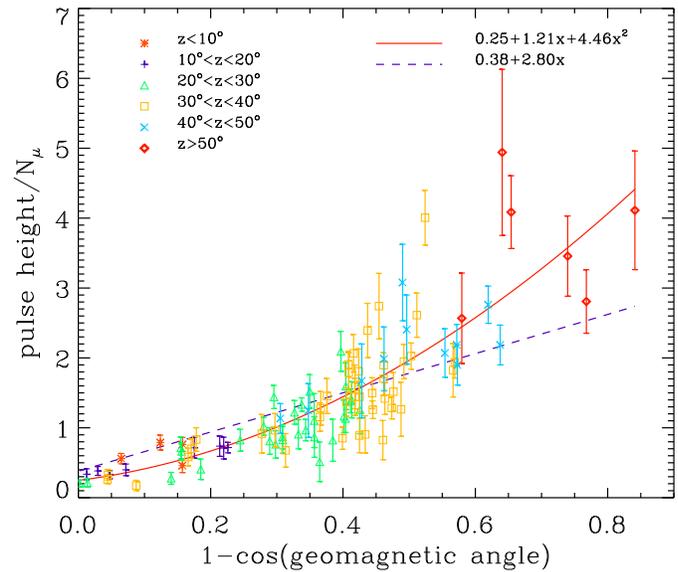}
\caption{\label{pulsegeo}Radio pulse normalized with the truncated muon 
number (and multiplied with 10$^6$) as a function 
of a geomagnetic angle for bright events \citep{2006PhDT........1H} and 
inclined events from this paper in the regime of 100\% radio detection.
The dashed line represents the linear fit for bright events with 
z$<$50$^{\circ}$,
and the solid line represents the quadratical fit for all events.}
\end{center}
\end{figure}

Figure \ref{zenitgeo} shows geomagnetic and zenith angles of inclined 
events triggered by the KASCADE detector in the year 2004.
Out of those, 28 events have accurately
reconstructed muon and electron numbers
by KASCADE (stars), and 21 were detected by LOPES-10 (rhombs).  
Note that there are 7 events with accurate KASCADE reconstruction that 
are detected by LOPES-10. Other events (21) reconstructed by 
KASCADE are not registered in the radio domain.
Also, there are 14 events 'seen' by LOPES-10, but with no reliable
particle numbers from KASCADE.
Finally, there are 7 events not accurately 
reconstructed by KASCADE and also not detected by LOPES-10 (triangles).

We can see that above a zenith angle of 60$^{\circ}$, muon and electron 
number sreconstructed by KASCADE are not reliable, 
due to the detector sensitivity decrease for large inclinations. 
On the other hand, the most inclined cosmic ray air 
shower (registered as highly energetical with a large muon density on the 
ground) detected in the radio domain has 
a zenith angle of almost 70$^{\circ}$.
Note that almost all events with a large geomagnetic angle 
(more than 60$^{\circ}$) are detected by LOPES-10.
This corresponds to a direction towards the north.
Events detected by LOPES with geomagnetic angles between 
30$^{\circ}$ and 40$^{\circ}$
are in the southern direction. 

The same can be seen in 
Fig. \ref{azimgeo}. It shows the geomagnetic angle and azimuth of 
49 inclined events. Symbols have the same meaning as in the Fig. 
\ref{zenitgeo}.
An azimuth of 0$^{\circ}$ corresponds to a direction towards the north. 
All radio detected 
events are $\pm$80$^{\circ}$ around the north and $\pm$40$^{\circ}$ 
around the south. Events from east and west directions are not detected. 
This may be due to two times larger antenna sensitivity in the north-south 
than in the east-west direction for inclinations larger than 
50$^{\circ}$ and/or due to the 
fact that showers coming from east or west are expected not to have
a significant emission in the east-west polarization component 
\citep{2005APh....24..116H},
which is the one that LOPES antennas measure.

There are more detected events coming from the north than from the 
south, 
which is probably related to the fact that cosmic rays coming from 
the north have the largest geomagnetic angles.  
Also, the tight correlation (two thin bands) shown in Fig. \ref{azimgeo} 
is
due to the fact that most events have zenith angles between 50$^{\circ}$ 
and 60$^{\circ}$, so the azimuth changes from 0$^{\circ}$ to 
360$^{\circ}$, 
whereas the zenith angle stays almost constant for those events.
The only two points out of this general trend are for zenith angles 
larger than 70$^{\circ}$.

Figure \ref{pulsegeo} shows events with zenith angles smaller 
than 
50$^{\circ}$ \citep{2006PhDT........1H} and inclined events from this 
paper with accurate KASCADE particle number reconstruction, which are in 
the region of 100\% radio detection considering 
their geomagnetic angles and truncated muon numbers. 
This can be explained as follows: The fraction of radio detected events 
within KASCADE triggered events increases both with rising muon number and with rising geomagnetic angle.
By defining a more strict cut on muon number and geomagnetic angle, 
1-cos(geomagnetic angle)$>$3.58-0.6log$(N_{\mu})$, as explained in detail by \citet{2006PhDT........1H}, it is
possible to get a selection where all events have been detected in the 
radio domain.
Figure \ref{pulsegeo} shows the correlation between the radio pulse 
normalized by the 
truncated muon number and the geomagnetic angle.
The normalized radio pulse height increases with the geomagnetic angle,
for highly inclined events also, which indicates the geomagnetic origin of
shower radio emission.
The value used for the radio pulse height in Fig. \ref{pulsegeo} 
is the height of a Gaussian fitted to cc-beam
and this value is affected by errors of the KASCADE data and 
statistic and systematic
errors connected with radio measurements \citep{2006PhDT........1H}.
We should point out that the gain of the LOPES antennas is calculated 
from the simulations
and that the real gain can differ from those values. The difference is the largest
for large zenith angles, due to the unknown reflections at the 
ground and the surroundings.
We conclude from the simulations that for zenith angles less than 
50$^{\circ}$, the effect 
on the pulse
height is less than 10\% \citep{2006PhDT........1H}. 
The corresponding estimated error for zenith angles larger than
50$^{\circ}$ is up to 50\%.
In Fig. \ref{pulsegeo}, beside other errors, 
we included 10\% gain error for events with z$<$50$^{\circ}$
and 50\% gain error for more inclined events. In this way, we took the 
largest possible
value of the error for highly inclined events into the account. 
We made a  linear fit for events z$<$50$^{\circ}$ and a quadratical fit
for all events. We can notice that even with a maximum gain error included, 
events with z$>$50$^{\circ}$ are located above the linear fit 
and match 
the quadratical dependence more closely.
However, we should keep in mind that a larger number of events 
with a large zenith angle might be necessary
for a final conclusions about this dependence,
and also, that LOPES-10 antennas
measure only east-west polarization and this may be the reason that 
events coming from the north have larger measured radio pulse heights.

\begin{figure}[h]
\begin{center}
\includegraphics*[width=\columnwidth,clip]{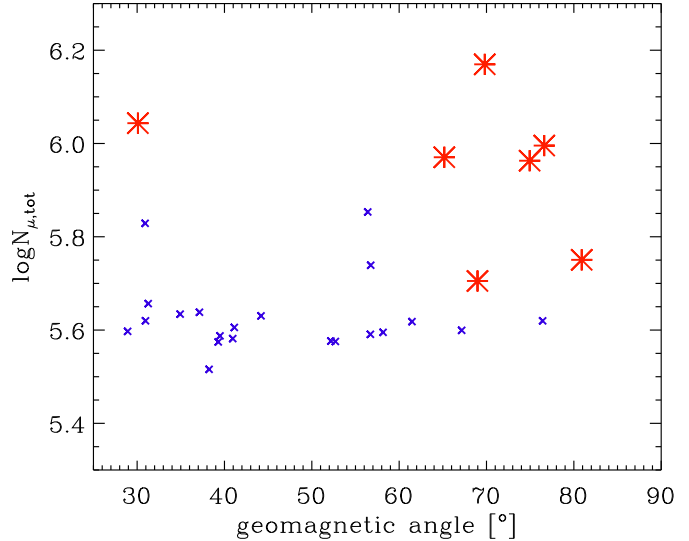}
\caption{\label{muogeo}Geomagnetic angle and the total number of muons (on the ground level) 
for inclined events from 2004 that have reliable KASCADE reconstruction 
of muon and electron number.
Big stars represent events detected by LOPES-10. 
Events with large geomagnetic angles and/or large muon numbers are 
favored for detection in the radio domain.}
\end{center}
\end{figure}

Figure \ref{muogeo} shows the geomagnetic angle and the total muon 
number (on ground level) of 28 inclined events with accurate KASCADE
electron and muon numbers.
Out of those, 7 events were detected by LOPES-10 (large stars).
Radio detection clearly favors cosmic ray air showers with a large 
muon number, which is in agreement with 
bright, nearly vertical events from
\citet{2005Natur.435..313F}. This is expected, since theory 
predicts that due to coherence \citep{2003A&A...412...19H},
the electric field strength scales approximately linearly with 
the number of electrons/positrons (times their track length) in the 
shower, which is 
roughly proportional to the number of muons in the case of 
the KASCADE detector.

\begin{figure*}[t]
\begin{center}
\includegraphics*[width=0.7\columnwidth,angle=270,clip]{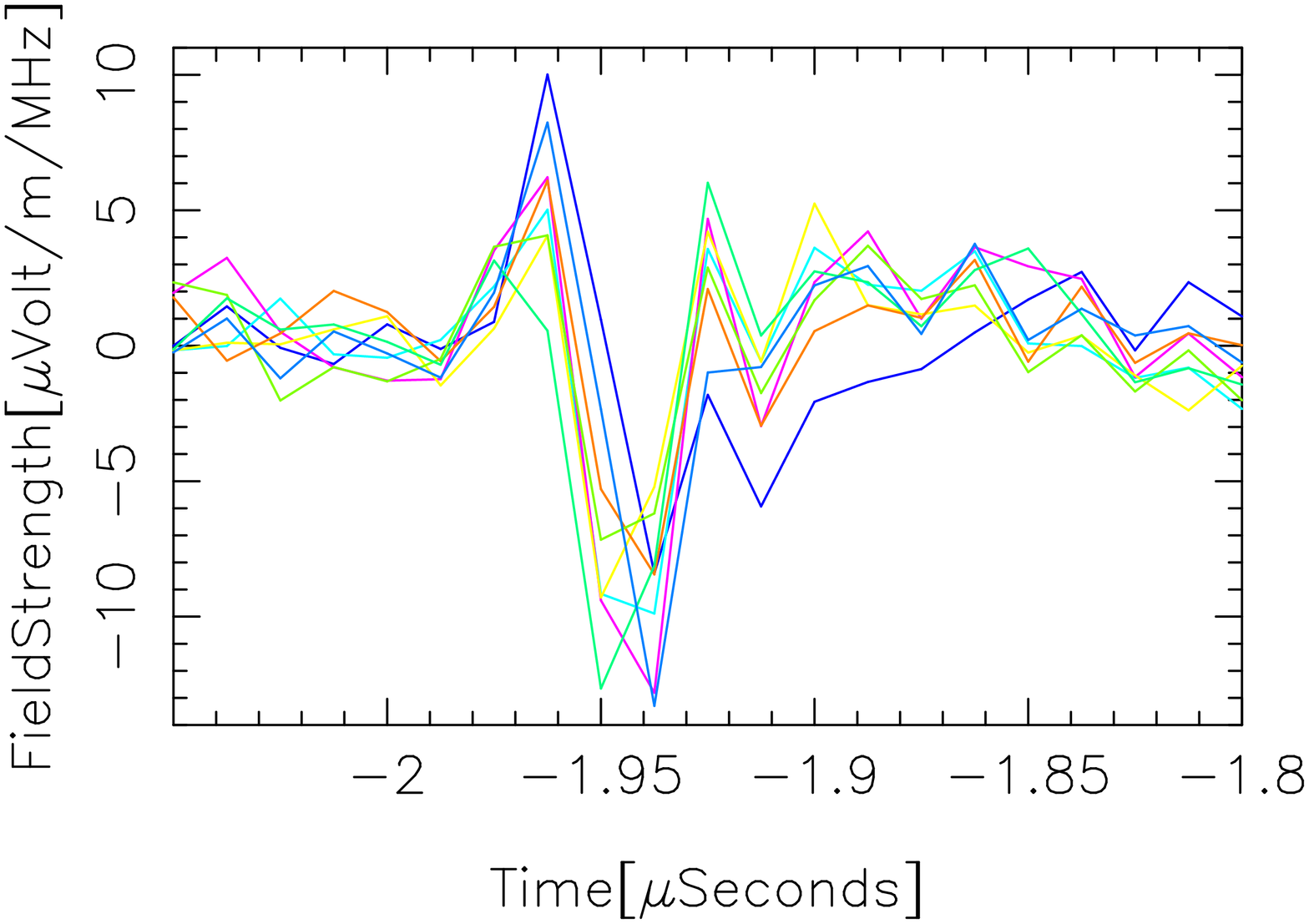}
\includegraphics*[width=0.7\columnwidth,angle=270,clip]{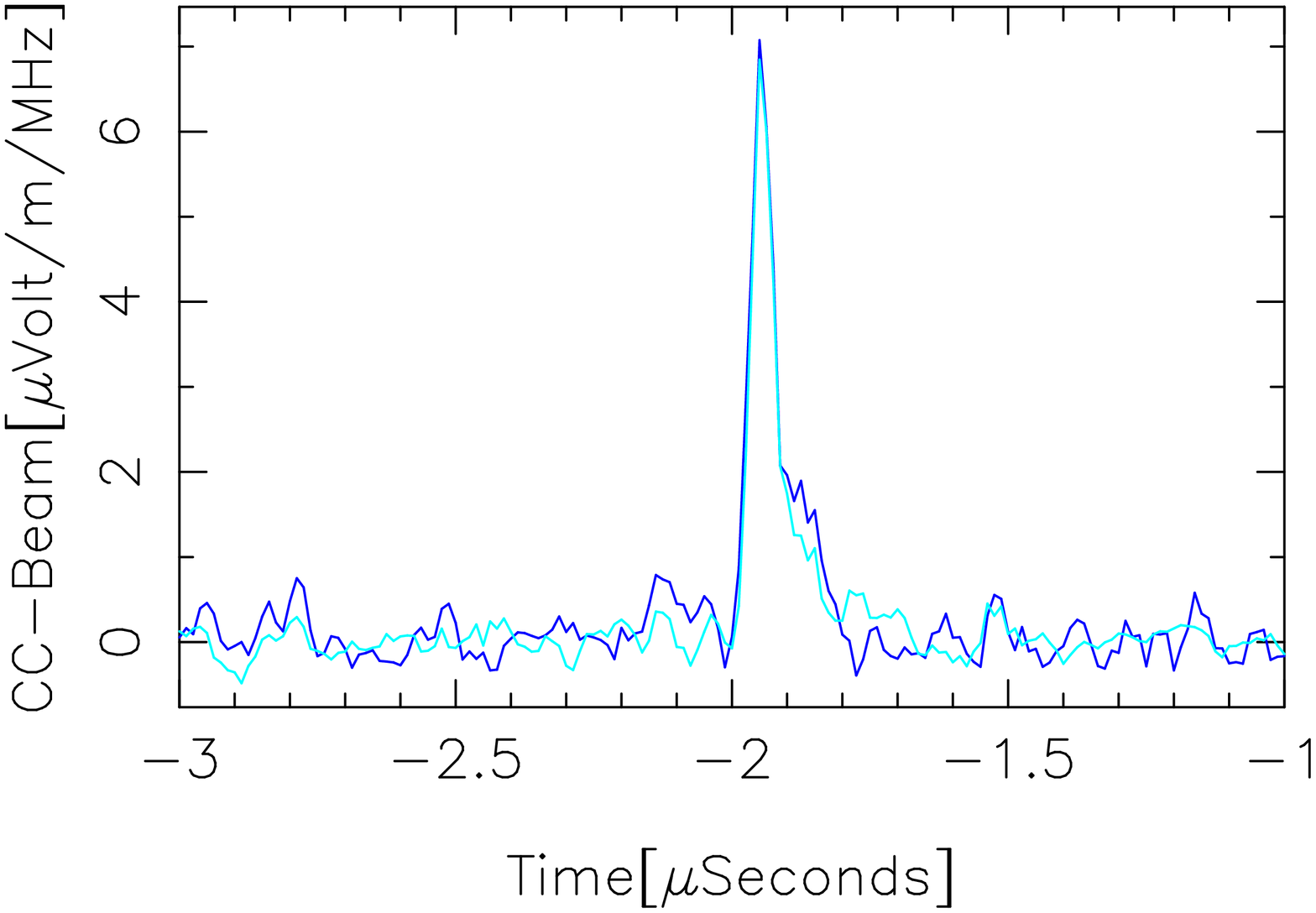}
\caption{\label{signal153-a}Left: Electric field as a function of time for 
LOPES-10 antennas for an event 
detected in January 2004 with a zenith angle of 77$^{\circ}$. The geomagnetic angle is 84$^{\circ}$. Right: Radio emission as a function of time after beam forming for the 
same event. The dark blue line represents cc-beam, the light blue line the 
power-beam. Note that the timescales on
the panels are different.}
\end{center}
\end{figure*}

\section{Events with a zenith angle larger than 70$^{\circ}$}

If we search through 86 KASCADE triggers for events with a zenith 
angle larger than 70$^{\circ}$, 
we find four events that are detected in the radio 
domain by LOPES-10. We do this without selecting events with a large muon 
density on the ground,
and that is why they do not show on plots in the previous section. 
This does not mean
that the number of particles in those showers is small, 
but due to the very large inclination, muon and electron numbers
are difficult to reconstruct with the KASCADE detector.

\begin{figure*}[t]
\begin{center}
\includegraphics*[width=0.7\columnwidth,angle=270,clip]{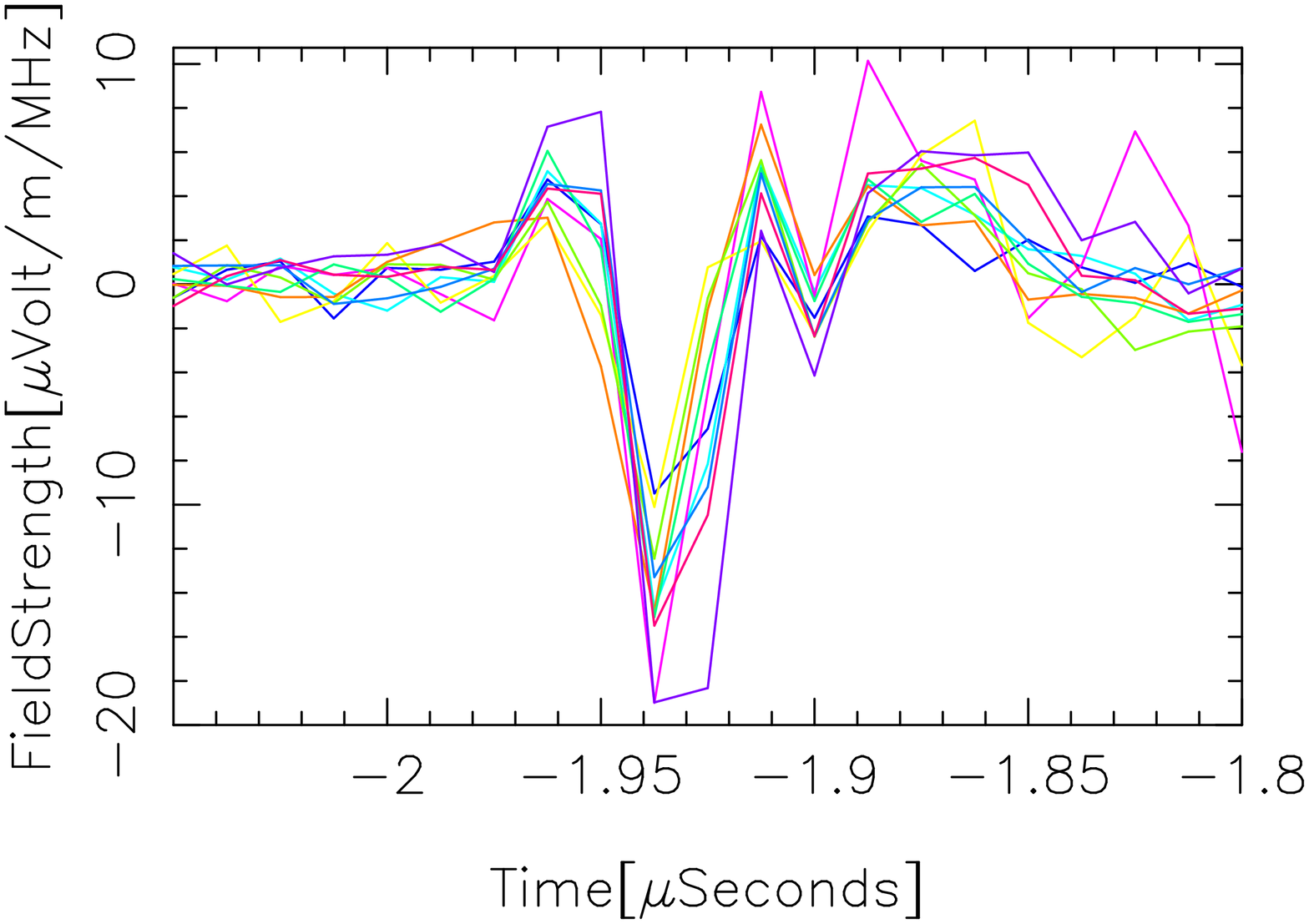}
\includegraphics*[width=0.7\columnwidth,angle=270,clip]{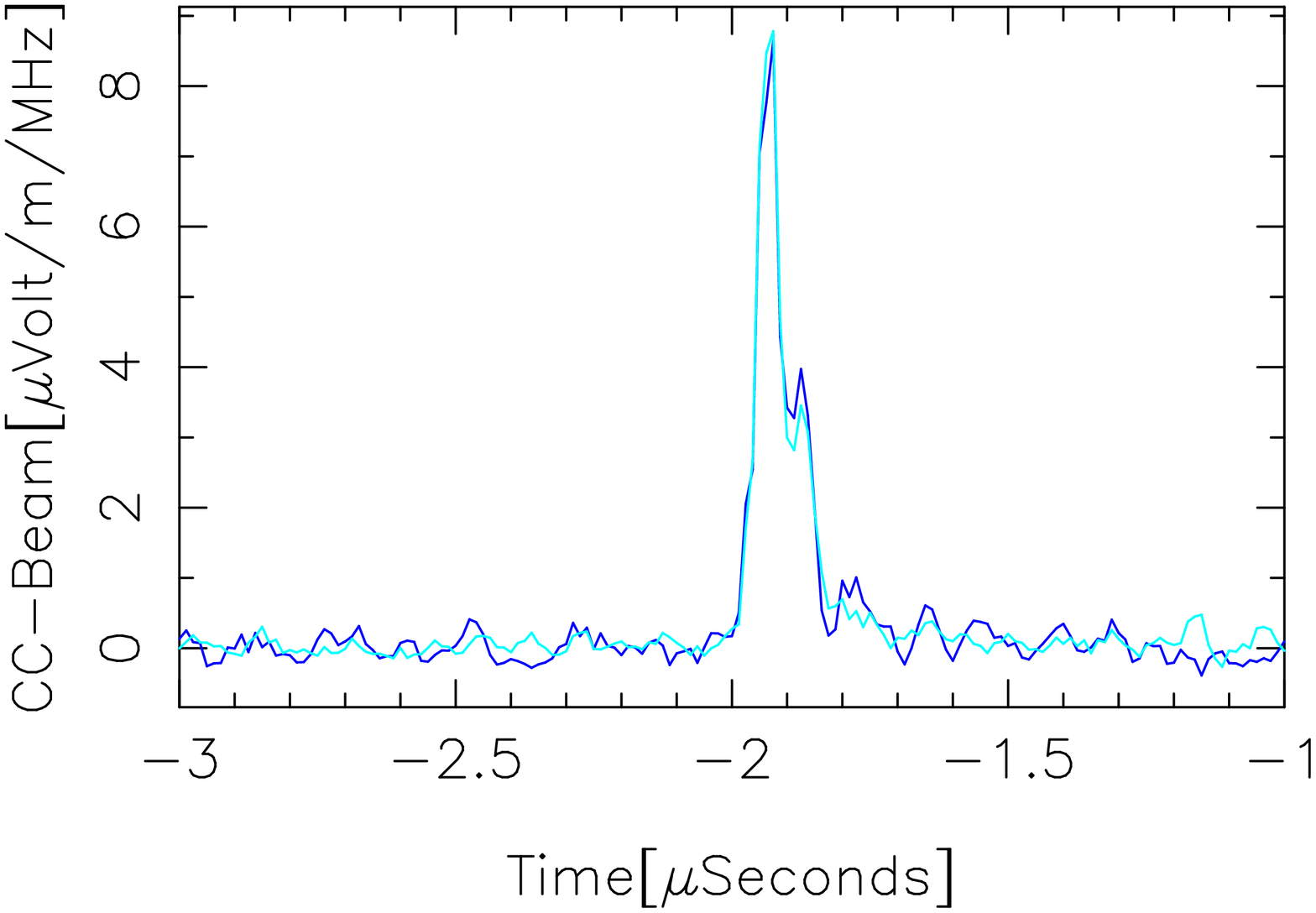}
\caption{\label{signal1397-a}Left: Electric field as a function of time 
for LOPES-10 antennas for an event
detected in July 2004 with a zenith angle of 73$^{\circ}$.
The geomagnetic angle is 88$^{\circ}$. Right: Radio emission as a
function of time after beam forming for the
same event. The dark blue line represents the resulting cc-beam, the
light blue line the power-beam.
Note that the timescales on the panels are different.}
\end{center}
\end{figure*}

We show two examples here.
The left panel of Fig. \ref{signal153-a} shows an event from January 2004. 
We checked and ruled out the possibility that
this event occurred in a thunderstorm enviroment \citep{buitink}.
This is the most inclined radio event detected by LOPES-10 during the year 
2004, 
with a zenith angle of 77$^{\circ}$ and an azimuth of 317$^{\circ}$.
The geomagnetic angle for this event is 84$^{\circ}$, 
which is again in agreement with the conclusion we have drawn 
earlier that events with a large geomagnetic angle are favored 
for radio detection. 
The signal is coherent with a very large field strength. 
Note that the noise after the peak is very weak, due to the low
noise from the photomultipliers, i.e., a small number of electrons 
that reach the ground, which is expected 
in the case of very inclined events.
A power-beam and a cc-beam formed pulses are shown in the right panel of 
Fig. \ref{signal153-a}. 
Another event with a zenith angle larger than 70$^{\circ}$ is shown 
in Fig. \ref{signal1397-a}.
Here, the thunderstorm connection is also ruled out.
This event is from July 2004 with a zenith angle 73$^{\circ}$, an
azimuth of 300$^{\circ}$,
and a geomagnetic angle of 88$^{\circ}$.

\section{Conclusions}

We have made a selection of events detected by LOPES-10 during
the year 2004 with a zenith angle larger than
50$^{\circ}$ and a large muon density on the ground, which corresponds
to a high energy of the primary particle (51 events).
Around 40\% of those events are detected in the radio domain,
and some of them with very high field strengths, like the example we 
presented in Sect. 3.
We also found four radio detected events with a 
zenith angle larger than 70$^{\circ}$ (Sect. 4). 

We found that inclined events that are preferred for radio detection are 
the ones with a large geomagnetic angle and/or a large
number of muons, which is proportional to the total number of
particles in a shower, i.e., the energy of the primary particle.
We also found that the radio pulse height
normalized with the truncated muon number increases with the
geomagnetic angle to larger zenith angles than considered in
\citet{2005Natur.435..313F}.
This implies that geomagnetic processes in cosmic ray air showers
are also dominant emission mechanism for highly inclined events.

For the particle detector KASCADE, the sensitivity drops
sharply towards the larger zenith angles.
However, this is not the case for radio detection.
Our results show that radio arrays can be used to detect very 
inclined cosmic ray air showers. We show that 
LOPES-10 can detect events up to a zenith angle of around 80$^{\circ}$.
This indicates that much larger radio detectors, if they are able to 
estimate the distance to the air shower maximum, might be used in the 
future as a possible detectors of neutrino induced showers with large 
inclinations initiated close to the ground.

However, for a possible future detection of neutrinos with a large 
radio array, like the LOFAR telescope, we
should keep in mind the following things: 

\begin{itemize}
\item
Detection efficiency of the
LOPES-10 radio array depends strongly on the geomagnetic angle and also, 
due to antenna sensitivity and polarization effects, on the azimuth angle. 
This makes certain directions favorable for radio detection and this 
is an important issue considering the very low flux of high energy 
neutrinos. 
For further understanding of geomagnetic angle dependence of a radio 
signal, it is necessary to understand better the emission processes in 
cosmic ray air 
showers. Azimuth angle detection efficiency dependence can possibly be 
solved by using antennas that measure both polarization directions.

\item
LOPES-10 data do not show radio detected events above 80$^{\circ}$, 
possibly
because of the missing triggers from the KASCADE array. In addition, the 
LOPES-10 antenna gain decreases towards the horizon and is not accurately 
known for large zenith angles. 
Knowing the exact effect of 
the radio wave reflections on the ground and surroundings
(now estimated to be up to 50\% on radio pulse height measured by 
LOPES-10 for highly inclined events) 
would greatly improve the estimate of the antenna gain at 
large zenith angles.
Another possible way to reduce this problem is to use 
an optimized antenna for measuring inclined showers, but
it is important to remember that 
one has to make a trade-off between antenna
sensitivity to the horizon and suppression of human-made radio emission, 
which usually comes from the horizon and decreases the signal-to-noise 
ratio even after beam forming.

\end{itemize}

\acknowledgements
This work is part of the research program of the Stichting voor 
Fundamenteel Onderzoek der Materie (FOM), which is financially supported
by the Nederlandse Organisatie voor Wetenschappelijk Onderzoek (NWO).
LOPES is supported by the German Federal Ministry of Education
and Research. The KASCADE-Grande experiment is supported by the German 
Federal Ministry of Education and Research, the MIUR of Italy, the Polish 
Ministry of Science and Higher Education, and the Romanian Ministry of 
Education and Research (grant CEEX 05-D11-79/2005). 
 
\bibliographystyle{aa}

\bibliography{5732lela}


\end{document}